\documentclass[twocolumn]{article}
\usepackage{amssymb,amsmath,bbm}
\usepackage{graphicx}
\newcommand{\ignore}[1]{{}}
\title{Isolation and connectivity in random geometric graphs with self-similar intensity measures}
\author{Carl P. Dettmann, University of Bristol, UK}
\date{\today}

\begin{document}

\twocolumn[
\begin{@twocolumnfalse}
\maketitle
\begin{abstract}
Random geometric graphs consist of randomly distributed nodes (points), with pairs of nodes within a given mutual distance linked.  In the usual model
the distribution of nodes is uniform on a square, and in the limit of infinitely many nodes and shrinking linking range, the number of isolated nodes
is Poisson distributed, and the probability of no isolated nodes is equal to the probability the whole graph is connected.  Here we examine these properties
for several self-similar node distributions, including smooth and fractal, uniform and nonuniform, and finitely ramified or otherwise.  We show that nonuniformity
can break the Poisson distribution property, but it strengthens the link between isolation and connectivity.  It also stretches out the connectivity transition.
Finite ramification is another mechanism for lack of connectivity.  The same considerations apply to fractal distributions as smooth, with some technical
differences in evaluation of the integrals and analytical arguments. 
\end{abstract}
\vspace*{10pt}
\end{@twocolumnfalse}
]

\section{Introduction}
Proposed by Gilbert~\cite{Gilbert61}, the random geometric graph (RGG)~\cite{Penrose03} was the original model of spatial
networks~\cite{Bart11}, and remains popular.  Nodes are distributed randomly in space, and links are formed between pairs of nodes with mutual distance
less than a threshold $r_0$.   This is a continuum model of percolation, since, when the density $\lambda$ increases, there is a sudden
transition to a state containing a large connected component~\cite{MR96}.  In a finite domain, typically a square or torus, under an appropriate
combined limit of $r_0\to 0$ and the expected number of nodes $\bar{N}\to\infty$, the probability that the network is a single connected
component has been widely studied.  In particular, it was shown by Penrose~\cite{Penrose97} and by Gupta and Kumar~\cite{GK99}
that in this limit, the number of isolated nodes (that is, nodes with no links) is Poisson distributed, and the probability of connectivity approaches
that of no isolated nodes.  As well as percolation and connectivity, other properties such as k-connectivity, clique number, chromatic number,
and Hamiltonicity are reviewed in Ref.~\cite{Walters11}. 

The soft random geometric graph (SRGG)~\cite{Penrose16,Krioukov16,MP15}
is a generalisation of the RGG with a second source of randomness: The links are formed independently with
a probability that is a given function $H(r)$ of the mutual distance $r$.  There are several communities using different names for this and closely
related models in the literature; see for example Waxman graphs~\cite{Waxman88}, continuum percolation~\cite{Penrose91,Alexander93}
random connection models~\cite{Bradonjic14,Mao17} and spatially embedded random networks~\cite{BDB07,PR17,HA17}.  Connection
functions can be constructed empirically with any spatial network for which the links can be assumed to be random~\cite{WDKD16};
several functions that have arisen from mathematical or physical arguments are given in Ref.~\cite{DG16}.

Most research on RGG and SRGG models has adopted the simplest model of a uniform distribution of nodes on a square or torus
domain. The main exceptions are as follows: Ref.~\cite{Penrose03} considers smooth densities that do not vanish supported on a cube,
and normally distributed densities. Ref.~\cite{HR05} considers more general smooth densities on $\mathbb{R}^2$ that decrease monotonically
outwards.  Ref.~\cite{GI10} considers (stretched) exponentially decaying density.  Ref.~\cite{ES15} considers rectangular domains, and
Refs.~\cite{CDG12b,DG16} consider arbitrary convex polygons and some domains that are polyhedral or have curved boundaries.   Poisson
distribution of isolated nodes was proved in Refs.~\cite{Penrose03,HR05,GI10} and the asymptotic equivalence with connectivity in
Ref.~\cite{Penrose03}; in none of this literature were exceptions to either statement observed or discussed.

One class of applications that has received substantial attention and for which physically motivated connection functions are available is that
of wireless mesh networks, in which devices (nodes) link to each other rather than to a central router or base station~\cite{CA11}.  Here, the
assumption of uniform density on a square or torus is often unrealistic.  One of the main sources of nonuniformity is node mobility.
In the random waypoint mobility model~\cite{BRS03,PDG16}, nodes move from one random point in the domain to another at a fixed or
random speed, then wait for a zero, fixed or random time.  In this model the average density of nodes is nonuniform, varying with position
in a smooth manner. If the wait time is zero, the average node density vanishes at the boundary, being roughly proportional to the distance
from a smooth edge. More sophisticated human mobility models are even more nonuniform, for example the popular self-similar least
action walk (SLAW) algorithm uses waypoints chosen from a fractal distribution~\cite{LHKRC09}.  

The purpose of this paper is to explore the world of node distributions beyond the existing literature, to include smooth densities that vanish
at the boundary, and fractal distributions.  In particular, we consider analytically and numerically whether  the results regarding the
distribution of isolated nodes and its link to connectivity remain true, in the limit of many nodes and for a small number.  Many interesting features can be
described using a relatively tractable class of distributions, the self-similar measures.  Extensions to other fractal measures are discussed
in Sec.~\ref{s:out}.

An important distinction throughout this paper is whether the distribution of nodes is ``almost uniform'' (AU).  We define a measure
$\mu$ on $\mathbb{R}^d$ (or a more general metric space)
as almost uniform if there exists a constant $ \varepsilon>0$ such that for all $r>0$ and ${\bf x},{\bf y}\in\mbox{supp}(\mu)$,
\begin{equation}\label{e:hom}
\mu(B({\bf x},r))\geq \varepsilon\mu(B({\bf y},r))\qquad 
\end{equation}
where $B({\bf x},r)$ is the ball of radius $r$ centred on ${\bf x}$.   Both the name AU and the concept are from Ref.~\cite{Studeny83} except that here the
${\bf x}$ and ${\bf y}$ are not arbitrary points in the space but belong to the support of the measure.
More discussion and context of this definition are found in Sec.~\ref{s:hom} below.

The structure of this paper is as follows: In Sec.~\ref{s:pre} we give the concepts needed later as found in the previous literature.  In Sec.~\ref{s:ex}
we describe the examples and a first numerical simulation.  Sec.~\ref{s:iso} investigates the distribution of isolated nodes, whilst Sec.~\ref{s:conn}
considers its link to connectivity.  Sec.~\ref{s:scale} makes some further observations about scaling the intensity by a contact factor, and Sec.~\ref{s:out}
concludes.

\section{Preliminaries}\label{s:pre}
\subsection{Quantifying non-Poissonness}
A random variable $X$ defined on the non-negative integers with probability mass function $\mathbb{P}(X=j)=P_j$ is Poisson distributed with mean $\mu$ if
\begin{equation}
P_k=\frac{\mu^k}{k!}e^{-\mu}
\end{equation}
To measure deviations from Poissonness, we consider factorial cumulants~\cite{Barbour87}, defined as
\begin{align}
q_n=&\left.\frac{d^n}{dt^n}\ln\mathbb{E}(t^X)\right|_{t=0}\\
=&n!\sum_{\{m_k\}:\sum km_k=n}(-1)^{\sum m_k -1}\nonumber\\
&\times\left(\sum m_k-1\right)!\prod_k\frac{1}{m_k!}\left(\frac{P_k}{P_0}\right)^{m_k}\qquad n>0\nonumber
\end{align}
The sum, obtained by writing the expectation explicitly in terms of the $P_j$ and expanding the logarithm,  is over integer sequences $\{m_k\}$ with $k\geq 1$ and $m_k\geq 0$.
It is easy to show using the first line that all $q_n$ for $n>1$ are zero for the Poisson distribution.
The first few factorial cumulants are
\begin{align}
q_0&=\ln P_0\nonumber\\
q_1&=\tilde{P}_1\nonumber\\
q_2&=2\tilde{P}_2-\tilde{P}_1^2\\
q_3&=6\tilde{P}_3-6\tilde{P}_2\tilde{P}_1+2\tilde{P}_1^3\nonumber\\
q_4&=24\tilde{P}_4-24\tilde{P}_3\tilde{P}_1-12\tilde{P}_2^2+24\tilde{P}_2\tilde{P}_1^2-6\tilde{P}_1^4\nonumber
\end{align}
where $\tilde{P}_j=P_j/P_0$.

\subsection{Poisson point processes}
We start with a $\sigma$-finite measure $\Lambda$ on $\mathbb{R}^d$.  Sometimes we have a smooth density $\lambda({\bf x})$ in which case
\begin{equation}
\Lambda(A)=\int_A\lambda({\bf x})d{\bf x}
\end{equation}
for measurable $A\subset \mathbb{R}^d$, however the only condition we will impose on $\Lambda$ is that it be nonatomic, that is, all sets of positive
measure contain a subset of smaller positive measure.
The Poisson point process (PPP) $\Phi$ with intensity measure $\Lambda$ is a random subset of $X$ defined by the following two properties~\cite{Haenggi12}:
\begin{enumerate}
\item For all measurable $A\subset \mathbb{R}^d$, $\Phi(A)$ is Poisson distributed with mean $\Lambda(A)$.
\item For all finite collections $\{A_i\}$ of mutually disjoint compact subsets of $\mathbb{R}^d$, $\Phi(A_i)$ are independent random variables.
\end{enumerate}
Here, $\Phi(A)=\sharp(\Phi\cap A)$, the number of points of $\Phi$  in $A$. If $\Lambda(A)$ is infinite, $\Phi(A)$ is almost surely infinite.
Unless otherwise stated, we will take $\Lambda(\mathbb{R}^d)=\bar{N}<\infty$ so that
$\bar{N}$ is the mean number of points, and $\Lambda_1=\Lambda/\bar{N}$ is a probability measure.  
We obtain a realisation of this process by first
choosing $N$ from a Poisson distribution with mean $\bar{N}$, then a binomial point process with $N$ points, that is, $N$ points chosen independently
with respect to $\Lambda_1$.

\subsection{Soft random geometric graphs}
The random geometric graph (RGG) model~\cite{Gilbert61,Walters11}  takes points according to a PPP and links all pairs with mutual distance less than a
threshold $r_0$.  The soft random geometric graph (SRGG) instead forms links independently with a probability $H(r)$ for points at mutual distance $r$.
Thus it has two sources of randomness, the point locations and the links.  One application is that of a wireless network with a mesh architecture, that is,
devices link to each other rather than to a central router or base station. Under the Rayleigh fading model (exponentially distributed channel gain,
neglecting interference), we find~\cite{HABDF09}
\begin{equation}\label{e:Rayleigh}
H(r)=\exp\left[-(r/r_0)^\eta\right]
\end{equation}
where $\eta>0$ is a constant called the path loss exponent.
Free propagation leads to $\eta=2$ (inverse square law); more cluttered environments have
$2<\eta<6$ and $\eta\to\infty$ recovers the RGG.  There are many more complicated fading models leading to $H(r)$ involving a number of special
functions~\cite{DG16}, however these can often be approximated by the above form~\cite{BDC13}.  Virtually all previous literature uses a uniform
(Lebesgue) intensity measure, perhaps on the finite domain.  Some smooth densities have also been considered, as noted in the introduction. 

\subsection{Isolation and connectivity}
The absence of isolated nodes is one piece of information obtainable from the degree distribution, and is an important characteristic of networks since it is necessary for connectivity.  In many cases, such as the original RGG in at least two dimensions, it is also sufficient with high probability~\cite{Walters11}.  If we denote by ${\cal N}_k$ the number of connected clusters of size $k$, then we have (using the same derivation as Eq.~5.1 of Ref.~\cite{Penrose15})
\begin{equation}\label{e:Eiso}
\mathbb{E}({\cal N}_1)=\int \exp\left[-\int H(|{\bf x}-{\bf x'}|)\Lambda(d{\bf x'})\right] \Lambda(d{\bf x})
\end{equation}
The exponential is simply the probability that a node at $x$ is isolated.  If $N_1$ is Poisson distributed, we have
\begin{equation}\label{e:Noiso}
\mathbb{P}({\cal N}_1=0)\to\exp\left[-\mathbb{E}({\cal N}_1)\right]
\end{equation}
Here, we consider a sequence of graphs with $\bar{N}\to\infty$ (but normalised measure $\Lambda_1$ fixed)
and suitable variation of the connection range $r_0$ (implicit in $H$)
so that the limit is positive and finite. Finally, if all other mechanisms for disconnection (for example, clusters of size two, or two large clusters)
become insignificant, the above expression is also the ``full connection'' probability:
\begin{equation}\label{e:Pfc}
\mathbb{P}({\cal N}_N=1)\to\exp\left[-\mathbb{E}({\cal N}_1)\right]
\end{equation}
in the same limit.  Both of the above statements hold for the original random geometric graph, where $H$ is a step function and the measure is uniform on a two or higher
dimensional cube~\cite{Walters11}.  Penrose has made significant progress generalising both of these statements to the SRGG with different connection functions~\cite{Penrose16},
and on the first condition where the measure can also be general~\cite{Penrose15}.  There are some conditions in the above papers that are not met by typical connection
functions such as those of Rayleigh fading, Eq.~(\ref{e:Rayleigh}), but it is likely that these are merely technical obstructions, and that all results hold for these connection functions.
In contrast, they rely nontrivially on some kind of uniformity condition, as in Eq.~(\ref{e:hom}) above or Eq.~(\ref{e:ehom}) below.  One of the main aims of this paper is to explore
the effects of non-uniformity.

\subsection{Self-similar measures}
Here, we assume that the intensity measure $\Lambda$ for the PPP is self-similar.  Self-similar sets and measures
are discussed at length in Ref.~\cite{Falconer14}, in particular in the chapter on multifractals.  We have a defining relation
\begin{equation}
\Lambda=\sum_{i=1}^m p_i \Lambda\circ T_i^{-1}
\end{equation}
where the $p_i$ are components of a nontrivial probability vector (values in $(0,1)$ that sum to unity), and the $T_i$ are a contracting similarity transformations
\begin{equation}
T_i({\bf x})=R_i{\bf x}+{\bf d}_i
\end{equation}
where $\frac{|R_i{\bf x}|}{|\bf x|}=r_i<1$ for all ${\bf x}\neq{\bf 0}$.  Hutchinson~\cite{Hutchinson79} showed that this defines a unique probability
measure with support given by the unique non-empty closed set $F$ satisfying
\begin{equation}
F=\cup_{i=1}^m T_i(F)
\end{equation}
The open set condition states that there is an open set $V$ such that $\cup_{i=1}^m T_i(V)\subseteq V$ and the $T_i(V)$ are mutually disjoint.  Under this condition
the Hausdorff dimension $D(F)$ is the unique positive solution of
\begin{equation}\label{e:ss}
\sum_{i=1}^mr_i^D=1
\end{equation}
More generally, we can consider self-affine measures, for which the transformation satisfies only $\frac{|R_i{\bf x}|}{|\bf x|}<1$ for all ${\bf x}\neq{\bf 0}$.
The dimension theory is much more interesting~\cite{Falconer13}, but from our point of view the analysis is similar.

Choosing a point at random with respect to this measure involves iterating the corresponding iterated function system (IFS).  Namely, start from an arbitrary initial point.
Then choose an $i\in\{1,\ldots,m\}$ with probability $p_i$ and apply $T_i$ to the point, repeating until the product of contraction ratios is smaller than the machine precision.  Likewise,
 it is straightforward to compute integrals over self-similar measures by iterating all combinations of the defining transformations:
\[ \int f({\bf x}) d\Lambda({\bf x})=\lim_{n\to\infty}\sum_{\{s_j\}}f(T_{s_1}\circ T_{s_2}\ldots T_{s_n}{\bf x}_0)\prod_{j=1}^np_{s_j} \]
for continuous function $f$ and arbitrary initial point ${\bf x}_0$.  The numerical computation is of course truncated to a finite depth $n$.  The numerical computations
are somewhat inefficient compared with techniques for standard integrals; a cycle-based approach~\cite{AAC90} may be more effective.  Analytical integrals over fractal 
self-similar measures are rarely available in closed form; polynomial integrands are an exception~\cite{DF93}.

\subsection{Almost uniformity}\label{s:hom}
An important concept in the theory of soft random geometric graphs is that of uniformity, sometimes used synonymously with homogeneity.  Penrose~\cite{Penrose15} defines a connection function
$H$ to be $\varepsilon$-homogeneous if
\begin{align}\label{e:ehom}
&\inf_{{\bf x}\in\mbox{supp}\Lambda}\int H(|{\bf x}-{\bf x'}|)\Lambda(d{\bf x'})\\
&\geq \varepsilon \sup_{{\bf x}\in\mbox{supp}\Lambda}\int H(|{\bf x}-{\bf x'}|)\Lambda(d{\bf x'})\nonumber
\end{align}
This is clearly a property of both the connection function and the measure $\Lambda$ used to define the PPP; here we are interested in the dependence on the measure.
For the Rayleigh fading model  Eq.~(\ref{e:Rayleigh}) with fixed connection range $r_0$
and measure with compact support there will always be some $\varepsilon>0$ for which this holds.  The question is whether it holds
uniformly as $r_0\to 0$.  If we must have $\varepsilon\to 0$ as $r_0\to 0$ for the hard connection (RGG) model $\eta=\infty$ we obtain exactly that
$\Lambda$ is non-AU as defined by Eq.~(\ref{e:hom}).  For all of the measures considered here, the definition of AU is equivalent if we replace
$\Lambda(B(x,r))$ by its equivalent for finite $\eta$, that is, $\int\exp[|{\bf x}-{\bf x}'|^\eta/r^\eta]\Lambda(d{\bf x}')$. 

We now discuss relations between the AU measures as defined in Eq.~(\ref{e:hom}) and the literature.
Requiring that the points $x$ and $y$ belong to the support of the measure means that, for example, the uniform
measure on the Sierpinski carpet satisfies our definition, but not that of Ref.~\cite{Studeny83} unless the space is taken to be the carpet itself, with metric induced from
its embedding into $\mathbb{R}^2$. The AU definition is also similar to the quasi-uniform measures of Ref.~\cite{JD08}, which are defined for symbol spaces.
A further connection is that an AU measure satisfies the condition for a doubling measure~\cite{Heinonen12}, since the support of the measure within a ball of radius
$2r$ may be covered by a bounded number of balls of radius $r$ centred at points in the support, where the bound depends only on the dimension $d$ of the ambient space.
Finally, if the $\varepsilon$ in the definition is replaced by unity, we obtain the widely studied uniformly distributed measures~\cite{KP02}.

For a general self-similar measure with the open set condition, it is easy to see that small regions of equal size will have similar measure (and hence satisfy the AU condition)
if and only if $p_i=r_i^D(F)$ (see Eq.~\ref{e:ss}).  The self-affine case is more subtle; see Ref.~\cite{FK11}.

Almost all existing results for (soft) random geometric graphs are for AU measures; exceptions are the smoothly decaying densities
discussed in the introduction.  Here we want to explicitly test the effects of non-uniformity (in the sense of AU).

\begin{table*}
\centerline{
\begin{tabular}{|c|c|c|c|c|c|c|}\hline
Name&$p_i$&$R_i$&${\bf d}_i$&Smooth?&Almost Uniform?&Finitely ramified?\\\hline
Uniform square&\multicolumn{3}{|c|}{Binomial square, $p=1/2$}&Y&Y&N\\\hline
$\lambda(x,y)=4xy$&1/16&$I/2$&$(0,0)$&Y&N&N\\
&1/8&$I/2$&$(1/2,0)$&&&\\
&1/16&$M_x/2$&$(1,0)$&&&\\
&1/8&$I/2$&$(0,1/2)$&&&\\
&1/4&$I/2$&$(1/2,1/2)$&&&\\
&1/8&$M_x/2$&$(1,1/2)$&&&\\
&1/16&$M_y/2$&$(0,1)$&&&\\
&1/8&$M_y/2$&$(1/2,1)$&&&\\
&1/16&$-I/2$&$(1,1)$&&&\\\hline
Binomial square&$p^2$&$I/2$&$(0,0)$&N&N&N\\
&$p(1-p)$&$I/2$&$(1/2,0)$&&&\\
&$p(1-p)$&$I/2$&$(0,1/2)$&&&\\
&$(1-p)^2$&$I/2$&$(1/2,1/2)$&&&\\\hline
Sierpinski triangle&$1/3$&$I/2$&$(0,0)$&N&Y&Y\\
$D=\ln 3/\ln 2$&$1/3$&$I/2$&$(1/2,0)$&&&\\
&$1/3$&$I/2$&$(1/4,\sqrt{3}/4)$&&&\\\hline
Sierpinski carpet&$1/8$&$I/3$&$(0,0)$&N&Y&N\\
$D=\ln 8/\ln 3$&$1/8$&$I/3$&$(1/3,0)$&&&\\
&$1/8$&$I/3$&$(2/3,0)$&&&\\
&$1/8$&$I/3$&$(0,1/3)$&&&\\
&$1/8$&$I/3$&$(2/3,1/3)$&&&\\
&$1/8$&$I/3$&$(0,2/3)$&&&\\
&$1/8$&$I/3$&$(1/3,2/3)$&&&\\
&$1/8$&$I/3$&$(2/3,2/3)$&&&\\\hline
\multicolumn{4}{|c|}{Power law, $\lambda(x)=cx^\alpha$, $x>0$ }&Y&N&Y\\\hline
\end{tabular}}
\caption[]{\label{t:trans} Defining transformations for the examples, using matrices $I=\mbox{diag}(1,1)$,
$M_x=\mbox{diag}(-1,1)$, $M_y=\mbox{diag}(1,-1)$.  The power law example is an infinite self-similar measure,
not defined from transformations.}
\end{table*}

\section{Examples}\label{s:ex}
The examples of self-similar measures we consider are defined in Tab.~\ref{t:trans} and (apart from the power law) are illustrated in
Fig.~\ref{f:i} below.  There are three characteristics that distinguish them, as expressed in Tab.~\ref{t:trans}, namely smoothness,
almost uniformity (as defined above), and whether the support is finitely ramified.

\begin{figure}
\centerline{\includegraphics[width=300pt]{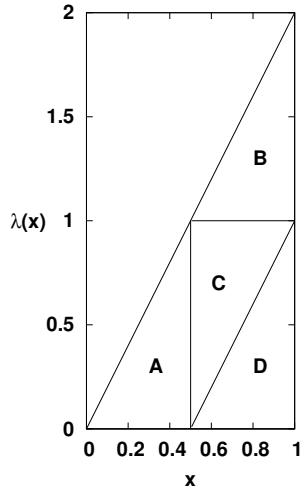}}
\caption{\label{f:4xy} Construction of a finite linear profile with density $\lambda(x)=2x$ as a self-similar measure.  It is equal to four copies of itself labelled $\{A,B,C,D\}$; $C$ has reversed orientation whilst $B$ and $D$ are identical and so can be combined.  Taking a product of this measure in the $x$ and $y$ directions gives the smooth $4xy$ construction defined in Tab.\protect\ref{t:trans}.} 
\end{figure}

With regard to smoothness,  note that self-similar measures need not be fractal; the examples include the uniform measure on the unit square
(the ``control'' example, extremely well studied in the literature) and the density $4xy$ on the unit square. The construction of the latter is
illustrated in Fig.~\ref{f:4xy}.

The 1D power law $\lambda(x)=cx^\alpha$ is self-similar with respect to dilations about the origin, but is
not defined by an IFS.  It is also an infinite measure.  For $\alpha\in[-1,0]$ it has infinitely many isolated nodes at large $x$, irrespective of $c$,
however for $\alpha<-1$ or $\alpha>0$ the expected number of isolated nodes is finite and hence interesting from our point of view.  Note that
the usual RGG limit of high density does not make sense here, since varying $c$ (the quantity corresponding to density) is exactly equivalent to
varying the connection range $r_0$: If we replace the connection range $r_0$ by $ar_0$ and $c$ by $a^{-\alpha-1}c$ for $a<1$, the system
exactly scales and the distribution of all ${\cal N}_k$ remain invariant. So, there is no sequence of graphs with which to
define the above limits; in this sense the system is always finite, or for $-1\leq\alpha\leq 0$, infinite but disconnected.
For simulations the domain is truncated at very small or large $x$ so as to give $\bar{N}=10^3c$.

With regard to almost uniformity, the $4xy$ and 1D power law (with $\alpha\neq 0$) are non-AU, as is the binomial square (with $p\neq 1/2$).

A finitely ramified set (usually in the context of fractals) is one for which large regions can be disconnected by the removal of a finite and bounded
number of points; here we see that the Sierpinski triangle satisfies that with a bound of 3, and that a 1D set (such as the power law example)
satisfies it trivially.  The importance of this condition is that it allows the
(soft) random geometric graph to easily split into large connected components, in contrast to measures such as the uniform square for which it
is known that isolated nodes are the dominant mechanism for disconnection.

\begin{figure*}
\centerline{\includegraphics[width=490pt]{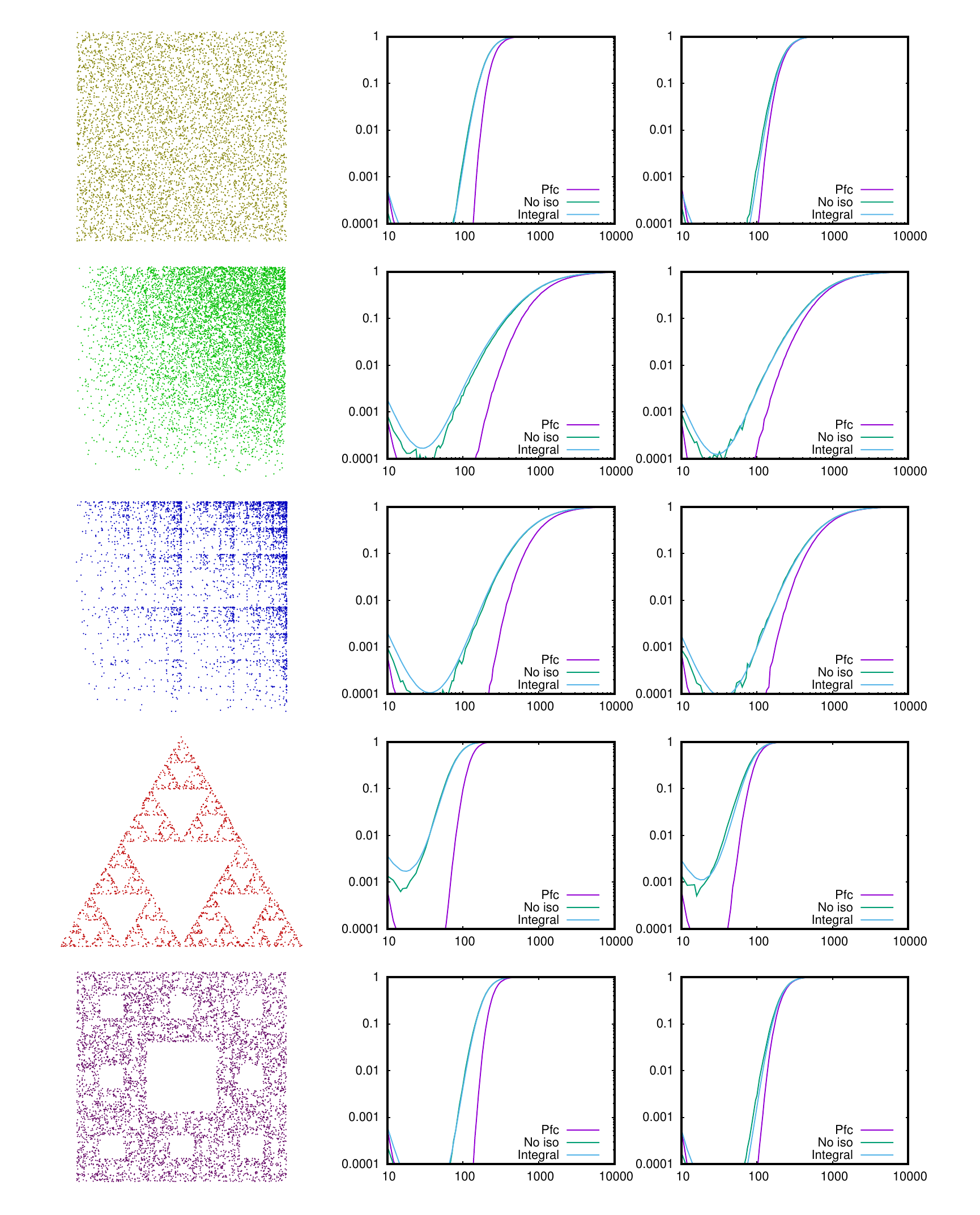}}
\caption{\label{f:i} Probability of no isolated nodes (integral and simulation) and of connectivity (simulation).
The horizontal axis is $\bar{N}$, the average number of nodes.
For this and the following plots, the left column is $\eta=\infty$ (RGG) and the right column is $\eta=2$ (SRGG).  The
connection range is $r_0=0.1$.  From top to bottom, the measures are: Uniform, $4xy$, binomial square with $p=3$,
Sierpinski triangle, Sierpinski carpet.}
\end{figure*}

Fig.~\ref{f:i} is a direct numerical test of both equations, (\ref{e:Noiso}) and (\ref{e:Pfc}), except for the 1D power example as discussed above.
The integral, performed directly by iterating the transformations, is compared
with a simulation of the network, and in particular the probability of no isolated nodes and of connectivity.  The computational time for the double integral is
roughly proportional to $m^{2\delta}$ where $\delta$ is the depth of the computation, that is, the number of iterations used in representing the measure.
The integral is performed with $\delta=5$ for the smooth $4xy$ ($m=9$) and Sierpinski carpet ($m=8$), and with $\delta=8$ for the uniform square
($m=4$), binomial
square ($m=4$) and Sierpinski triangle ($m=3$).  The smooth $4xy$ measure required $\delta=6$ for the $\eta=\infty$ case for convergence, although in this
case other integration methods are clearly available.  It is seen that Eq.~(\ref{e:Noiso}) is well satisfied but Eq.~(\ref{e:Pfc}) is not, showing that there are
other mechanisms for connectivity than isolated nodes.  The connection range $r_0$ is of course not particularly close to zero, so this does not contradict the
outcomes expected in the asymptotic limit discussed in the sections below.  We now consider distribution of isolated nodes and connectivity properties separately, both theoretically and with further tailored numerical simulations.

\section{Poisson distribution of isolated nodes}\label{s:iso}
\subsection{Focus: $4xy$ model}
One existing and useful approach is that of Hsing and Rootzen~\cite{HR05}.  Their Theorem 1 gives sufficient conditions for the number of isolated nodes in a RGG
to be Poisson distributed in a limit of many nodes where the expected number of isolated nodes is positive and finite, so $r_0$ decreases with $\bar{N}$.
This does not assume uniformity or smoothness of the measure, though non-smooth measures are not discussed explicitly in that paper.  In rough terms,
the assumption is that it is possible to cover almost all of the domain with blocks separated by twice the connection range so that the number of isolated nodes in each block is
independent.  In addition, the expected number of isolated nodes in each block is small, and the ratio of the expected number of pairs of isolated nodes with distance in
$[r_0,2r_0]$ (hence, correlated) to the number of isolated nodes is also small.  It is easy to see that for AU measures, including the relevant
fractal examples, splitting the domain into blocks much larger than the connection range but much smaller than the system size satisfies these conditions.

Let us now consider the non-AU examples.  For the $4xy$ model, we note that the density of isolated nodes is given from Eq.~(\ref{e:Eiso}) by
\begin{equation}\label{e:lami}
\lambda_i({\bf x})=\bar{N}4xy\exp\left[-\int\int H(|{\bf x}-{\bf x'}|)\bar{N}4x'y'dx'dy'\right]
\end{equation}
In the bulk, so when $x$ and $y$ are further from the edge than $r_0$, and using the RGG (step function) connection function, the integral is
\begin{equation}
\int\int H(|{\bf x}-{\bf x'}|)\bar{N}4x'y'dx'dy'=\bar{N}4xy\pi r_0^2
\end{equation}
Integrating from $r_0$ to $\infty$ (since isolated nodes will occur only near the origin) we conclude
\begin{equation}
\mathbb{E}(N_1)>\frac{1}{4\bar{N}\pi r_0^4}\left[e^{-4\bar{N}\pi r_0^4}+\int_{4\bar{N}\pi r_0^4}^\infty t^{-1}e^{-t}dt\right]
\end{equation}
and thus for $\mathbb{E}(N_1)$ to be finite, we must have $\bar{N}r_0^4\not\to 0$.  At the edge, for $y=0$ and $x>r_0$ we have
\begin{equation}
\int\int H(|{\bf x}-{\bf x'}|)\bar{N}4x'y'dx'dy'=\bar{N}\frac{8}{3}x\pi r_0^3
\end{equation}
so that $\lambda_i$ decays exponentially with $x$.  As a result, if $\bar{N}r_0^4\not\to 0$, all isolated nodes are of order $r_0$ from the origin with high probability
and the conditions of the Hsing-Rootzen theorem fail.  Thus, we do not expect the number of isolated nodes to be Poisson distributed.

The above argument applies to many other smooth densities that vanish at the boundary, including a corner where the density vanishes as a power of distance.
Exceptions, for which isolated nodes are Poisson distributed, include that of a symmetric disk (so there is no corner at which isolated nodes concentrate), and very slow
vanishing at a corner such as $\lambda(r)=-1/\ln r$ with radial coordinate $r$.

\subsection{Focus: Binomial square}
For the binomial square it is challenging to do an accurate calculation, but we can see that for a small connection range $r_0$, we have reached the iteration of the transformation
at level $k\approx -\frac{\ln r_0}{\ln 2}$.  The corner at this level has measure $p^{2k}$, and its vicinity has some expected number of isolated nodes depending on $\bar{N}$.
There are also $2k$ regions with measure $p^{2k-1}(1-p)$ and $2k(2k-1)/2$ regions with measure $p^{2k-2}(1-p)^2$, and so on. In a limit of large $\bar{N}$ and small
$r_0$, hence increasing $k$, any appreciable probability of an isolated node at the corner will be overwhelmed by the increasing number of regions with comparable probabilities
of isolated nodes.  Thus, if the total expected number of isolated nodes approaches a finite limit,
we expect to construct blocks, none of which have a significant probability of an isolated node.  So, the Hsing-Rootzen theorem appears to hold,
and the number of isolated nodes should be Poisson in the limit.  Note that $k$ increases only logarithmically with $\bar{N}$, so convergence may be slow.  Also, this is only
a sketch; careful estimates would be needed for a rigorous proof.  It remains open how to construct non-AU fractal distributions for which the Poisson property does
not hold.

For the SRGG case (soft connections), the Hsing-Rootzen approach is not longer valid since it is not possible to construct blocks where the isolated nodes are truly independent;
a Chen-Stein approach as used in other literature is likely needed.  Random connections do however reduce the correlations, so it is an interesting open question as to whether
the soft connection model restores Poissonness in the cases where it is broken (for example the $4xy$ model), or for what connection functions.  The soft connection model
introduces more randomness, so is likely to make the distribution of the number of isolated nodes closer to Poisson.  Note, however, that whilst the Hsing-Rootzen approach
does not strictly hold for soft connection models, the question as to whether there are an unbounded number of isolated nodes is a global question and so unaffected by short
ranged connection functions.  So, we do not expect that the soft model behaves differently to the hard model in the limit.   We now turn to numerical simulations.

\begin{figure*}
\centerline{\includegraphics[width=350pt]{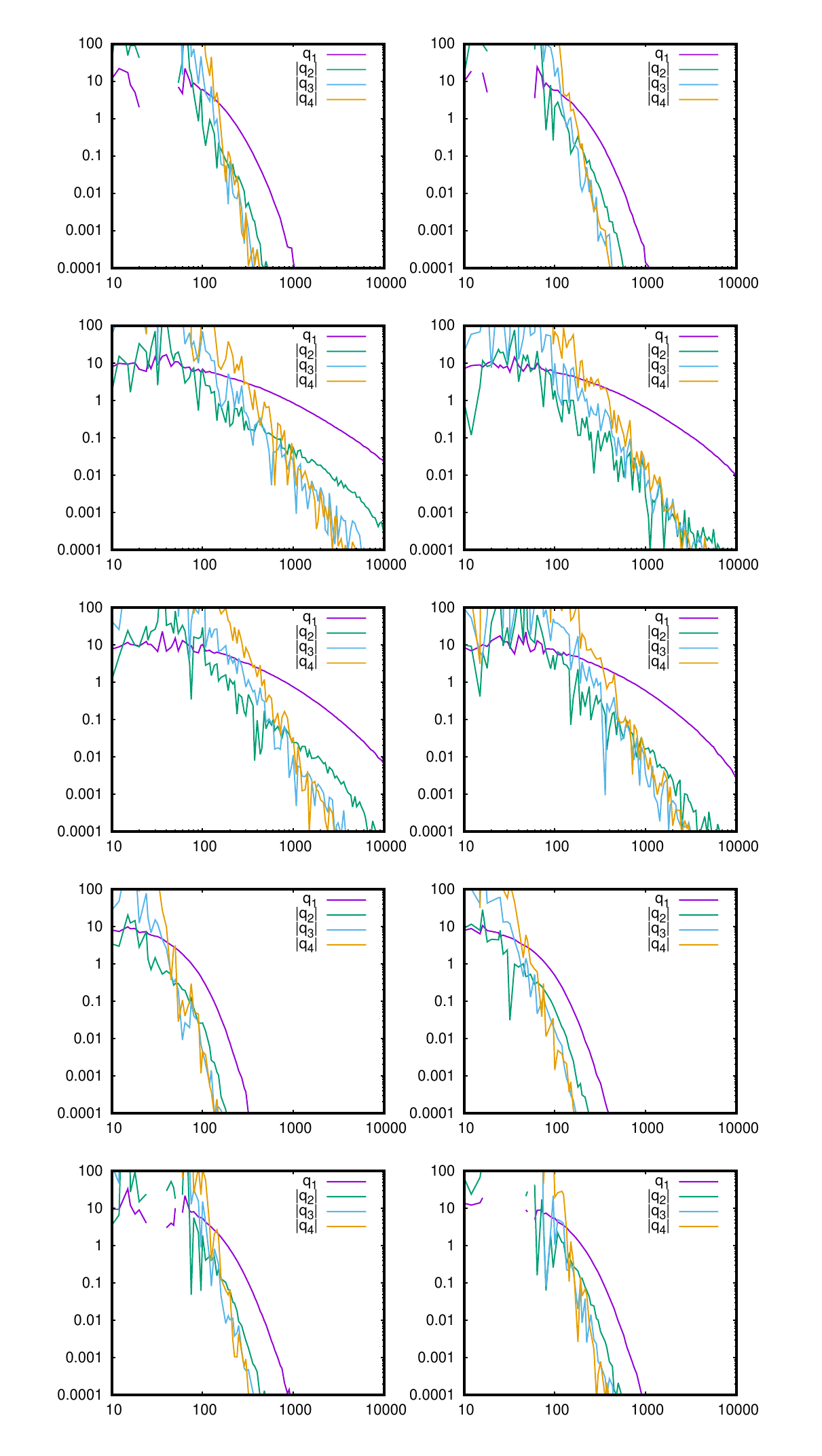}}
\caption{\label{f:p} Factorial cumulants in the distribution of the number of isolated nodes; models and parameters as in Fig.~\protect\ref{f:i}}
\end{figure*}

\begin{figure*}
\centerline{\includegraphics[width=350pt]{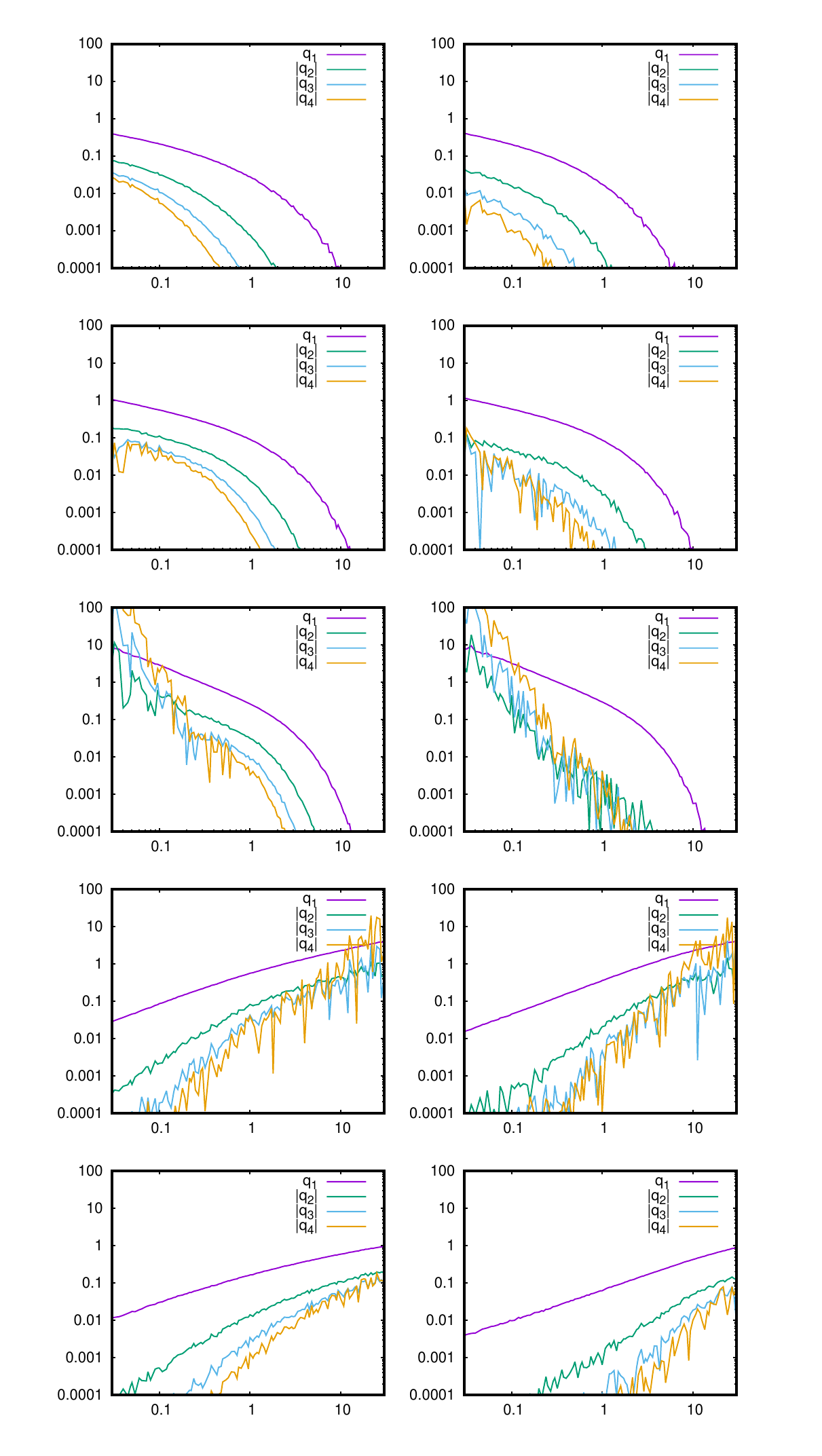}}
\caption{\label{f:pp} Factorial cumulants in the distribution of the number of isolated nodes; 1D power law, $\alpha=3,2,1,-2,-3$ from top to bottom, left
is $\eta=\infty$, right is $\eta=2$.}
\end{figure*}

\subsection{Numerical results}
Although Eq.~(\ref{e:Noiso}) is well satisfied in Fig.~\ref{f:i}, there must be a deviation from the Poisson distribution for the number of isolated nodes at some level;
this is tested in Figs.~\ref{f:p} and~\ref{f:pp},
which plot the factorial cumulants against $\bar{N}$ for the 2D models and against $c$ for the 1D models.  For large numbers of nodes, we see in Fig.~\ref{f:p} that $q_1$
is well above the other curves as expected.  We see that $|q_2|$ decreases similarly in all cases, except the non-AU cases ($4xy$ and the binomial square)
at $\eta=2$ in which it is
smaller for small $q_1$.  Thus, the non-AU models have most extreme behaviour, most deviations from Poisson distribution for $\eta=\infty$ (RGG)
as predicted above, but also much smaller deviations than the AU models for $\eta=2$.  This difference between $\eta=\infty$ and the still short-ranged
model $\eta=2$ is surprising, and conflicts with the rough argument given above, perhaps because we are unable to attain the limit $\bar{N}\to\infty$ numerically.
The 1D power law results in Fig.~\ref{f:pp} are consistent with these observations, noting that $|\alpha|$ gives an indication of the level of nonuniformity.

A related question to the distribution of the number of isolated nodes is whether the isolated nodes form a Poisson point process.  We know that the
isolated nodes beyond a distance of $2r_0$ are independent (almost independent for the SRGG model), so the deviations from a PPP are of the form
we already discussed here, namely that there are correlations for distances below $2r_0$ and that if the isolated nodes are concentrated in a finite
number of special points (for example corners), these correlations will invalidate the Poisson distribution (and hence PPP) of isolated nodes.  Likewise,
if they are not so concentrated, the independence will ensure that apart from short ranged correlations, the distribution of isolated nodes will look
like a PPP.  Because there are many possible ways of measuring deviation from a PPP, and because the results are expected to duplicate what we
find for the number of isolated nodes, we have not investigated the spatial distribution of the isolated nodes.

\section{From isolation to connectivity}\label{s:conn}
\subsection{Focus: 1D power}
Now, we consider Eq.~(\ref{e:Pfc}) in more detail, namely the statement that in the limit, the presence of isolated nodes is the only non-negligible mechanism for lack of
connectivity of the full network.  Alternative mechanisms may be roughly categorised as small or large components.  In the former case, clusters of two or three nodes
become as significant as isolated nodes.  In the latter, the network splits into two or more large pieces.  In this section, we consider 2-clusters in a simple model without the
AU property, and large clusters for a finitely ramified fractal.

We consider the effects of the non-AU property in the simplest possible model, the 1D power law example.
The connection function is the unit disk, that is $r_0=1$. Note that one-dimensional RGG models allow the network to break into large pieces.  In this calculation we are not
interested in this effect, only the formation of small clusters.  We have (see Eq.~\ref{e:Eiso})
\begin{align}
\lambda_i(x)&=\lambda(x)\exp\left[-\int\mathbbm{1}_{|x-y|<1}\lambda(y)dy\right]\nonumber\\
&\approx cx^\alpha\exp\left[-c\frac{(x+1)^{\alpha+1}}{\alpha+1}\right]
\end{align}
where the second line is exact for $x<1$ (most relevant in the high density limit we are considering).  Integrating, using a Laplace expansion for small $x$, we find
\begin{align}
\mathbb{E}(N_1)&=\int_0^\infty\lambda_i(x)dx\nonumber\\
&\approx \Gamma(\alpha+1) c^{-\alpha}\exp\left[-\frac{c}{\alpha+1}\right]
\end{align}
Generalising the formula for the expected number of isolated nodes (see for example Ref.~\cite{LNS16}), we consider the density of 2-clusters, that is, pairs of nodes that are
linked to each other but to no other nodes in the network:
\begin{align}
\lambda_2(x,y)=&\lambda(x)\lambda(y)\mathbbm{1}_{|x-y|<1}\nonumber\\
&\times\exp\left[-\int\left(1-\mathbbm{1}_{|x-z|>1}\mathbbm{1}_{|y-z|>1}\right)\lambda(z)dz\right]\nonumber\\
\approx& c^2x^\alpha y^\alpha \exp\left[-c\frac{(y+1)^{\alpha+1}}{\alpha+1}\right]
\end{align}
where we have assumed without loss of generality that $y\geq x$ and the approximation is valid for small $x$ and $y$, specifically $x<1$ and $y-x<1$.  Integrating,
again using Laplace expansions, we find
\begin{align}
\mathbb{E}(N_2)&=\int_0^\infty\int_0^y\lambda_2(x,y)dxdy\nonumber\\
&\approx \Gamma(2\alpha+2)\frac{c^{-2\alpha}}{\alpha+1}\exp\left[-\frac{c}{\alpha+1}\right]
\end{align}
And so
\begin{equation}
\frac{\mathbb{E}(N_2)}{\mathbb{E}(N_1)}\approx \frac{\Gamma(2\alpha+2)}{\Gamma(\alpha+2)}c^{-\alpha}
\end{equation}
Thus the number of 2-clusters decays more rapidly with density for greater nonuniformity parameter $\alpha$.  In this (high density) limit, nonuniformity
improves the link between isolation and connectivity.

But in terms of the usual limit, in which the number of nodes increases and the connection range decreases so that the expected number of
isolated nodes reaches a finite limit, the system scales as discussed in Sec.~\ref{s:ex}.
In particular, the expected number of 2-clusters is independent of the scale, and hence also non-negligible.

\subsection{Focus: Sierpinski triangle}
The other example we consider in this context is the Sierpinski triangle.  This set is finitely ramified, which means that many regions can be
disconnected from the rest of the set by the removal of a finite number of points.  Here, any of the small triangles may be isolated by the
removal of at most three points.  Thus if the regions of these points are empty in the PPP, it is quite feasible to expect very large connected
components.  It is very difficult to do precise calculations, but there are some bounds and general arguments through which we can gain
insight into this effect.

The set has three outer corners.  There are also infinitely many points which are vertices of small triangles which we call inner corners.
A triangle containing an outer corner will be called an outer triangle, whilst others will be called inner triangles.
The measure on a small triangle at level $k$ is $\bar{N}3^{-k}$.  For $r_0=2^{-k}$, it is sufficient to vacate two such triangles adjacent to
an inner corner to prevent links near the corner; this has probability $\exp(-2\bar{N}3^{-k})$.  In order to isolate an inner triangle, we
need to vacate its three corners, giving a probability $\exp(-6\bar{N}3^{-k})$ whilst to isolate an outer triangle we vacate only its two
inner corners, with probability $\exp(-4\bar{N}3^{-k})$.

A triangle we isolate in this way must have level $j<k-1$ in order not to be completely emptied.  At level $j$, there are $3^j-3$ inner
triangles which are not empty with probability $1-\exp\left[-\bar{N}(3^{-j}-3^{-k+1})\right]$ and $3$ outer triangles which are not
empty with probability $1-\exp\left[-\bar{N}(3^{-j}-2\times 3^{-k})\right]$.  For $j<k-2$ the emptiness probabilities are close to unity,
and the isolation probabilities do not depend on $j$, so it is more likely to find relatively small isolated inner clusters (that is, high $j$)
simply because there are more of them.

Suppose there is a nontrivial probability that at least one of the outer triangles is isolated.  Thus $\left[1-\exp(-4\bar{N}3^{-k})\right]^{3k-6}$
is of order unity, and hence $4\bar{N}3^{-k}\approx \frac{1}{\ln k}$.  Changing the prefactor from $4$ to $6$ for the inner triangles,
we see that the number of such triangles, which is exponential in $k$, dominates, and many of the inner triangles will be isolated.  Thus
close to the connectivity transition, only the inner triangles are relevant.  A similar argument shows that at this transition (in contrast
to higher densities) the isolation of a node near the corner is not relevant compared to the bulk.  This is completely analogous to the
usual result in random geometric graphs on the uniform square, for which the corner nodes are not relevant at the connectivity transition
(again, in contrast to much higher densities).

Finally, we note that a node at an inner corner (for simplicity) is isolated with probability $\exp(-2\bar{N}3^{-k})$, the cube root of the
probability that an inner triangle is isolated.  There are $(3^{k+1}-3)/2$ inner corners, which is the same order as the number of inner
triangles.  Thus, if there is a nontrivial probability of an isolated node, there will also be a nontrivial probability (perhaps much closer to
zero) that an inner triangle will be isolated.  Hence large clusters, specifically at relatively small inner corners, are relevant in the limit
where the expected number of isolated nodes is finite.  There are of course many approximations in this argument, but none of them
are likely to affect the conclusion. 

\begin{figure*}
\centerline{\includegraphics[width=350pt]{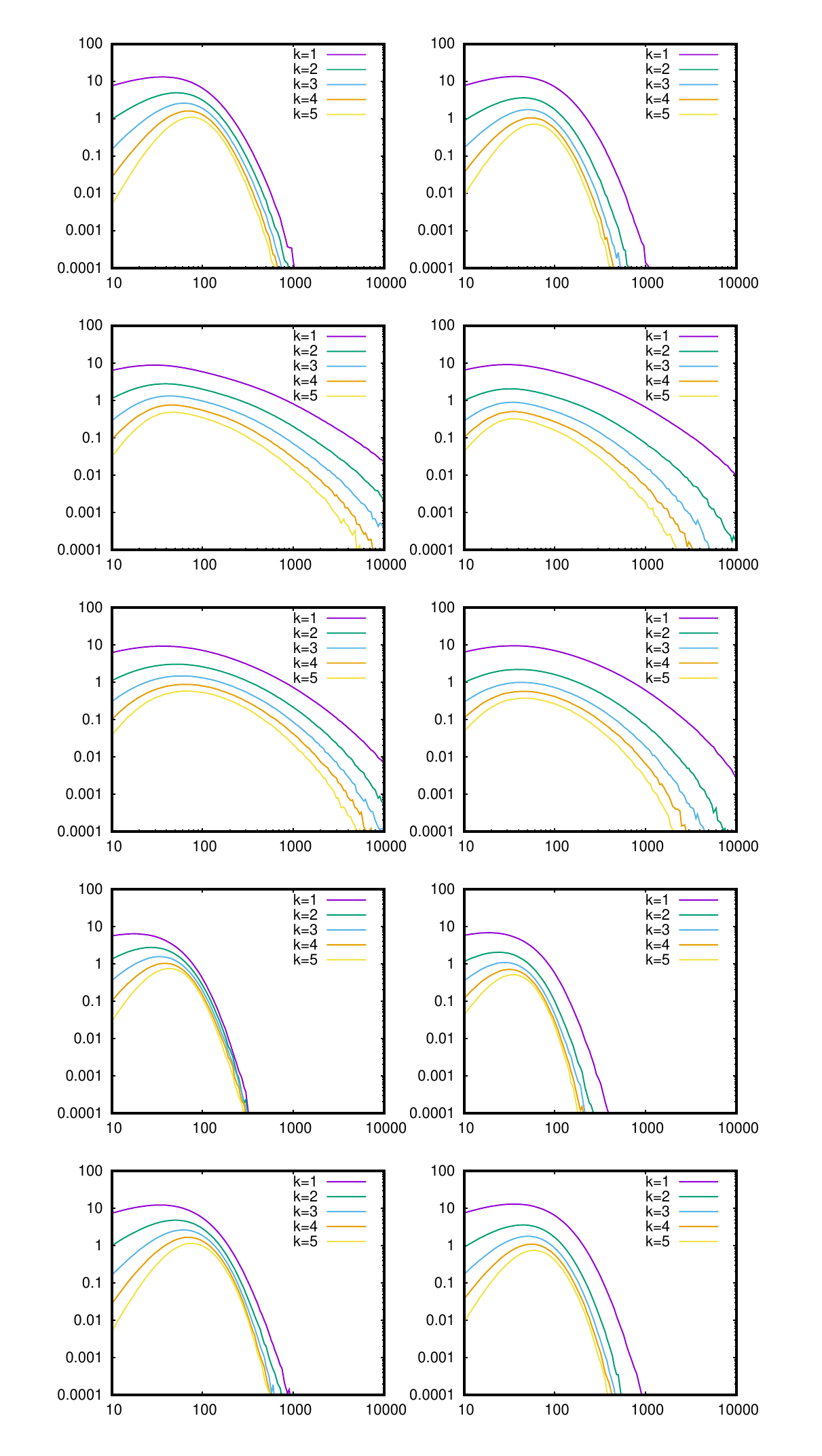}}
\caption{\label{f:c} Expected number of clusters of size $k$; models and parameters as in Fig.~\protect\ref{f:i}}
\end{figure*}

\begin{figure*}
\centerline{\includegraphics[width=350pt]{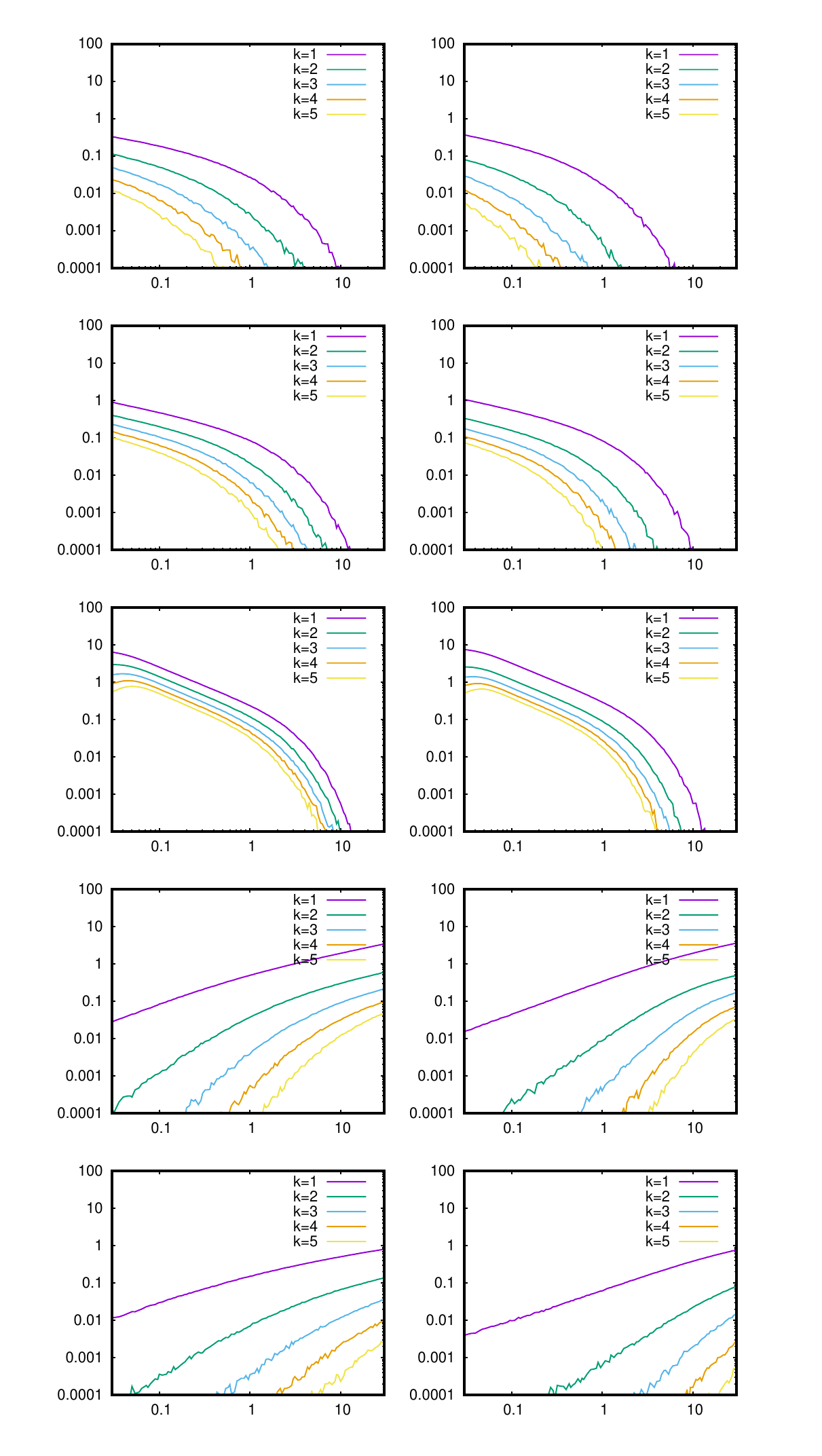}}
\caption{\label{f:cp} Expected number of clusters of size $k$; 1D power law, as in Fig.~\protect\ref{f:pp}}
\end{figure*}

\subsection{Numerical results}
The simulations in Figs.~\ref{f:c} and~\ref{f:cp} are concerned with the formation of small clusters, which break the link between isolation and lack of
connectivity, as observed above.  Again we see that isolated nodes are more common than larger clusters, but there are significant differences between the
various models.  Soft connections reduce the number of larger clusters in all cases, whilst for the original RGG model ($\eta=\infty$) we see that the number of
larger clusters decreases in the following order: Sierpinski triangle, uniform, Sierpinski carpet, binomial square, $4xy$.  Again, the non-AU property improves the
link between isolated nodes and connectivity.  The Sierpinski triangle is expected to have larger clusters in any case as discussed above, although
note that $r_0=0.1$ corresponds to $k\approx 3$ which is far from the asymptotic behaviour.  The 1D power law has many large clusters for the same reason,
mostly at $\alpha=1$ where it is most uniform.

We conclude this section with a further remark about the general context of connectivity and isolated nodes.  One popular model in the literature is
that of RGG on the hyperbolic plane~\cite{KPKVB10}.  Whilst the measure and distance functions, and hence the RGG model are naturally defined, this geometry is 
homogeneous but not scale invariant, so that the concept of self-similarity does not make sense.   Its structure is similar to that of a Cayley tree,
so that removing a finite number of nodes splits the RGG into several large components.  Hence lack of connectivity is again not necessarily due
to isolated nodes.

\section{Scaling the intensity by a constant factor}\label{s:scale}

\begin{figure}
\centerline{\includegraphics[width=250pt]{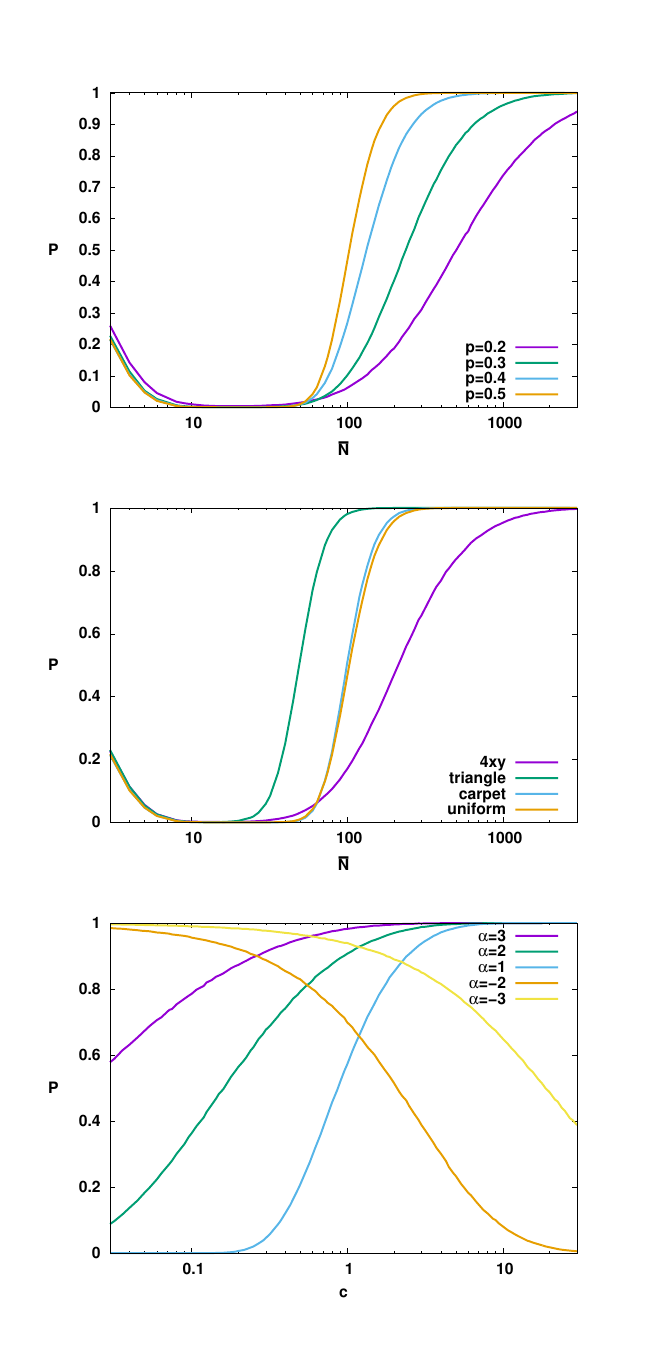}}
\caption{\label{f:stretch} Probability of connectivity as a function of number of nodes for the binomial square (upper plot), 1D power (lower plot) and other models
(middle plot).  Note that $p=0.5$ in the upper plot is the same as uniform in the middle plot.  Here, $\eta=2$; the RGG model
$\eta=\infty$ gives almost identical results.  The connection range is $r_0=1$ for the 1D power and $r_0=0.15$ otherwise.}
\end{figure}

We can also consider the probability of connection (and/or that there are no isolated nodes; the effects are similar) as a function of the number
of nodes for models with differing nonuniformity, see the top two panels of Fig.~\ref{f:stretch}.   
All the results are consistent with nonuniformity broadening the connectivity transition.  This occurs because a very high probability of connectivity requires that even the most sparse regions of the measure are reasonably covered with nodes.  But at smaller numbers of nodes, these regions are often vacant, and so do not contain the isolated nodes that would block connectivity.  These numerical results clearly show that the AU models,
whether smooth or fractal, have a similar width of transition, whilst the non-AU models have a much broader transition.

Actually, the story is more interesting than this.  It is easy to see that multiplying the density $\lambda(x)=e^x$ (which is not self-similar) by a constant is equivalent to translating it.  Considered as a RGG on the whole real line, it has a nontrivial connection probability, since the nodes at large positive $x$ are connected with very high probability, and there are no nodes at large negative $x$ to disconnect, also with very high probability.  So, multiplying this intensity measure by a constant leaves the connection probability invariant: There is no connectivity transition.  Results for this model may be obtained from Refs.~\cite{GIM08,GI10}.

Finally, if we take the power law density $\lambda(x)=cx^\alpha$ for $\alpha<-1$ then for small values of $c$ there are probably no nodes for $x>1$ and the infinite number of nodes in the unit interval are all connected.  For larger $c$, nodes extend to large $x$ with a gradually decreasing density and many of them will be isolated (and there will also be large gaps that prevent connectivity).  Thus, increasing the density reduces the probability of both connection, and of having no isolated nodes.  This is depicted in the lower panel of Fig.~\ref{f:stretch}.

\section{Outlook}\label{s:out}
We have seen both similarities and differences between uniform RGG and SRGG models and those with self-similar measures, both smooth and fractal, AU and non-AU and finitely ramified or otherwise.  Whether isolated nodes are Poisson distributed depends on whether they are concentrated in a corner or similar small region of the fractal.  Connectivity may be broken by small or large clusters as well as isolated nodes, particularly in the finitely ramified case. Strong nonuniformity can reverse the dependence of both of these on the intensity, making connectivity more likely at lower intensities.  The soft connection function tends to randomise both properties, but may not lead to qualitative differences.  As with the uniform case, finite systems may be far from the limiting behaviour.

The examples considered here have only scratched the surface of what is possible with self-similar measures, let alone self-affine measures, and
statistically self-similar measures.  The number of isolated nodes is only one of the simplest local graph properties; there are many others of
interest including the whole degree distribution and assortativity.  Connectivity is only one of the simplest global graph properties; there are
many others of interest including betweenness centrality and spectrum of the adjacency matrix.

From a practical point of view, the broadening of the connectivity transition in non-AU networks means that it is not cost effective to add nodes until the connection probability is
very close to unity; other means of ensuring connectivity such as adding them in specific locations, or of not requiring
connectivity, as in delay tolerant networks, are likely to be needed.     Of course, unless they are designed specifically in this manner, the intensity measure of real networks is not exactly self-similar.  The big challenge is the development of accurate models of complex environments.

\section*{Acknowledgements}
The author would like to thank J. Coon, E. Crane, O. Georgiou, J. Harrison, T. Jordan, G. Last, J. Mackay, M. Penrose and M. Wilkinson for helpful discussions.
This work was supported by the EPSRC grant number EP/N002458/1 for the project Spatially Embedded Networks. 

\bibliographystyle{custom3}
\bibliography{fractal,wireless}

\end{document}